\pdfoutput=1
\documentclass[cits]{JINST}
\usepackage{graphicx}
\usepackage{textcomp} 
\usepackage{booktabs} 
\usepackage{amsmath,amsfonts,amssymb,mathrsfs}
\title{The design and performance of an improved target for MICE}

\author{C.N.~Booth\thanks{Corresponding author}, P.~Hodgson, J.~Langlands, 
E.~Overton, M.~Robinson, P.J.~Smith\\
   Department of Physics and Astronomy, University of Sheffield, Sheffield S3 7RH, UK
   \email{C.Booth@sheffield.ac.uk}
  } 
\author{G.~Barber, K.R.~Long\\
  Department of Physics, Blackett Laboratory, Imperial College
    London, Exhibition Road, London SW7 2AZ, UK
  } 
\author{B.~Shepherd\\
    STFC Daresbury Laboratory, Daresbury, Cheshire WA4 4AD, UK
  } 
\author{E.~Capocci, C.~MacWaters, J.~Tarrant\\
  STFC Rutherford Appleton Laboratory, Chilton, Didcot, 
    Oxfordshire, OX11 0QX, UK
  } 
\abstract{
The linear motor driving the target for the Muon Ionisation Cooling 
Experiment has been redesigned to improve its reliability and performance.
A new coil-winding technique is described which produces better magnetic 
alignment and improves heat transport out of the windings.
Improved field-mapping has allowed the more precise construction to be demonstrated,
and an enhanced controller exploits the full features of the hardware, enabling
increased acceleration and precision.
The new user interface is described and analysis of performance data 
to monitor friction is shown to allow
quality control of bearings and a measure of the ageing of targets during
use.
}


\preprint{MICE-PUB-BEAM-480
}

\begin{document}
%
%
%
%
%
\section{Introduction}
\label{Sect:Intro}

%

The international Muon Ionisation Cooling Experiment, MICE, will provide an
engineering demonstration of a novel technique to reduce the phase space
of a muon beam, while furnishing the facility to study a variety of
materials and options for investigating the cooling process in 
detail \cite{MICE}.
The motivation for such a study derives from plans for future muon storage
rings as sources of high intensity neutrino beams at the Neutrino Factory
\cite{NF} and high energy lepton-antilepton colliders \cite{MC}.

MICE is hosted at the 800\,MeV proton synchrotron ISIS \cite{ISIS} at the
Rutherford-Appleton Laboratory \cite{RAL}.
A custom beam-line has been constructed, supplying muons of momentum between
140\,MeV/c and 240\,MeV/c \cite{beamline}.
The muons arise from the decay of pions, which themselves derive from 
interactions between the accelerated proton beam and a solid titanium
target.
ISIS operates at 50\,Hz and the target is dynamic, intercepting roughly 
one proton pulse a second, while being magnetically levitated out of the beam
between designated pulses.
High acceleration and precision of control are required in order to intercept
a pulse at the end of its acceleration cycle and be clear of the beam envelope
10\,ms later when the next pulse is injected.

The previous target \cite{PaperI} operated successfully on ISIS from 2011,
performing over half a million actuations during three years.
However, variability in performance between different examples of the target
mechanism, and the need to generate the absolute minimum of abraded material
(which could enter the accelerator vacuum space)
due to wear have led to a 
programme of improvements which are reported in this publication.

Under test, some target drives showed a degradation of performance after 
hundreds of thousands of actuations, while others (including the device installed
on the accelerator) performed perfectly for many millions of cycles.
It has been suggested that poorer performance could be due to off-axis
forces, due to misalignments between the stator coils in the body of the
drive and permanent magnets mounted on the titanium shaft the tip of which 
constituted the target.
The previous target mechanism employed a stator containing 24 pre-wound
and potted coils which were then assembled on the tube forming the
vacuum barrier and down which the shaft passed \cite{PaperI}.
Section \ref{Sect:Stator} describes a new construction technique which
allows the coils to be wound and potted in situ on a special tube which
also acts as a former.
This not only allows greater alignment precision; it also enables tighter 
magnetic coupling between the coils and permanent magnets.
This in turn allows higher acceleration to be achieved while reducing the
coil currents and associated ohmic heating, as discussed in section 
\ref{Sect:Performance}.
The redesign of the coils also allows improved water cooling, details of
which are given below.

Section \ref{Sect:Controller} describes improvements to the
control circuitry.
Our previous publication \cite{PaperI} explained how the stator coils are 
connected in three banks,
with two being powered at any one time; the new controller allows currents
to flow through all three banks simultaneously when this provides a 
greater driving force on the target.
The user interface to the target controller has been upgraded to 
allow greater functionality and easier operation by both experts and other
users; this is documented in section \ref{Sect:GUI}.
The new target has been extensively tested and its performance is summarised 
in section \ref{Sect:Performance}.
Finally, a summary is provided in section \ref{Sect:Summary}.

%
\section{Stator}
\label{Sect:Stator}
During the operation of the original ``T Series'' of targets, it was found that 
stator-to-stator performance varied considerably. 
Target T1, for example, performed well while subsequent models were less 
reliable. 
The acceleration dropped off over the life of the target and dust was created 
at the bearing positions. 
At first this was assumed to be due to a combination of bearing/shaft 
tolerances, bearing material and surface finish of the shaft. 
Over the testing period all these aspects were investigated and major 
improvements were made, but after these developments it was still apparent that 
T1 performed better that any other target.
One possibility considered was that because the stator construction consisted 
of a series of 24 individually wound coils assembled on a stainless steel tube 
the coils may not share the same magnetic axis; furthermore they may not be on the 
mechanical axis defined by the bearings. 
This was subsequently investigated, as described in section \ref{Sect:Stator:Field}, and 
offsets were found to be present both for the coil-to-coil magnetic axis 
alignment and the mechanical axis alignment. 
In view of the success of the bearing system in T1 and the fact that the body 
was designed to fit with the target support frame and the control systems,
it was decided to redesign the internal stator assembly whilst 
keeping the existing stator body. 
The revised design is presented below.

%
\subsection{Hardware design and manufacturing}
\label{Sect:Stator:Design}
To improve the coil-to-coil alignment, a one-piece mandrel was manufactured.
This allowed each coil to be wound in place, and as each coil was wound using 
the same set-up then it was hoped that they would all share a common magnetic
axis. 
This created two difficulties that needed addressing.
The individual coils must be wound side by side without damaging the preceding 
coil and whilst keeping the pitching exact.
The winding must also be carried out in such a way as to leave both tails of 
the coil accessible for connection.

As the stator was being redesigned the opportunity was also taken to reduce 
the radial clearance between the inner bore of the coils and the outer diameter 
of the permanent magnets mounted on the moving shaft, to improve the 
performance. 
Improvements were also made to the cooling.

\begin{figure}
\begin{center}
\includegraphics[width=0.9\textwidth]{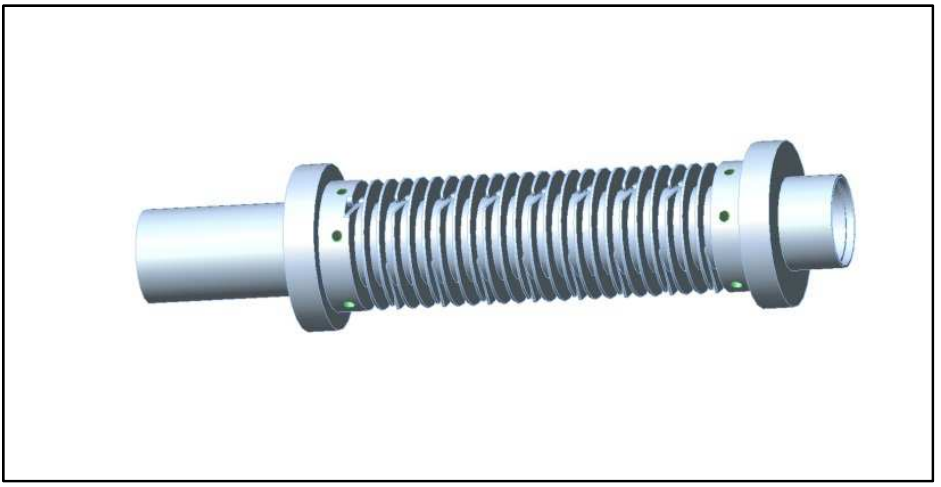}\\
\end{center}
  \caption{The mandrel, showing fins to separate the coils and slots for
    wire connections.}
  \label{fig:stator:bobbin}
\end{figure}

The resulting design is illustrated in figure \ref{fig:stator:bobbin}.
A stator mandrel was produced with dividing fins between each winding to ensure 
the correct pitch was maintained. 
Into each fin a slot was spark-eroded to allow the first power tail to pass to 
the bottom of the coil without encroaching on the width of the coil. 
A second slot was cut in three of the fins to allow thermocouples to be
embedded in the stator to closely monitor the temperature inside the assembly. 
These fins also give a heat path from the bottom of the coil to the outside 
diameter where the cooling jacket is positioned.

The mandrel was produced at AC Precision, Standlake and took several trials 
before the final procedure for manufacture was determined. 
This was due mainly to the fact that when such thin fins were machined they had 
a tendency to distort, which made the winding of the coil very difficult. 
The slots were then spark eroded at Oxford University Physics department 
workshop. 
Once the mandrel had been procured, it was surveyed for compliance with 
the specified dimensions and wound with the 24 individual coils 
at Rutherford-Appleton Laboratory. 
The wire used was 500\,$\mu$m square-section copper wire varnished for insulation. 
This was purchased from MWS Wire Industries, Westlake Village CA USA. 
The wire was ordered ``layer wound'' (i.e. with no twist on the bobbin), which 
was important for the winding process. 
Each coil consisted of 5 turns per layer with 10 layers making up the
complete winding.
(The final layer had 4 turns, making a total of 49 turns per coil.)
This resulted in coils which were slightly proud (by about 0.25\,mm)
of the mandrel, so that when subsequently a cooling jacket was fitted it 
rested firmly on the coils to ensure a good thermal contact.
When individual coils were complete, they were fitted with a temporary
plastic protection shell, to ensure no damage was done while winding
the rest of the stator.
After each coil was completed, and again when the complete assembly had been
wound, electrical tests were performed as described in section \ref{Sect:Stator:QC}.
Only when all tests were satisfied did construction
proceed to the next stage.

\begin{figure}
\begin{center}
\includegraphics[width=0.5\textwidth]{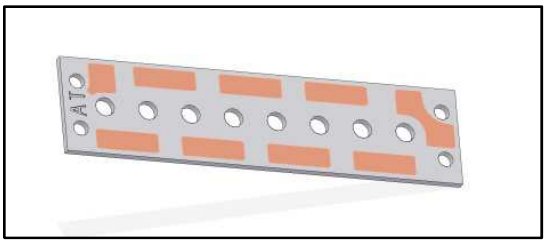}\\
\end{center}
  \caption{The printed circuit board used for power connections.}
  \label{fig:stator:pcb}
\end{figure}

%
As described in \cite{PaperI}, the 24 coils are wired in three phases, 
A B \& C in a repeating pattern. 
Each coil in a phase is connected in series, with one end of the phase 
connected to a power lead and the other connected to a common star-point. 
The phases are powered in pairs by the controller in the appropriate sequence 
to produce the required motion of the target shaft. 
To facilitate the correct wiring of the coils, printed circuit boards, as 
illustrated in Figure \ref{fig:stator:pcb},
were produced at Imperial College HEP electronics lab.
These boards were attached to the mandrel and then the coils and power leads 
soldered to them. 
A connection was also made directly to the star connector to allow individual 
coils to be powered in isolation for tests and field mapping.
The thermocouple devices were also bonded to the assembly. 
Once more, electrical tests were carried out. 

\begin{figure}
\begin{center}
\includegraphics[width=0.7\textwidth]{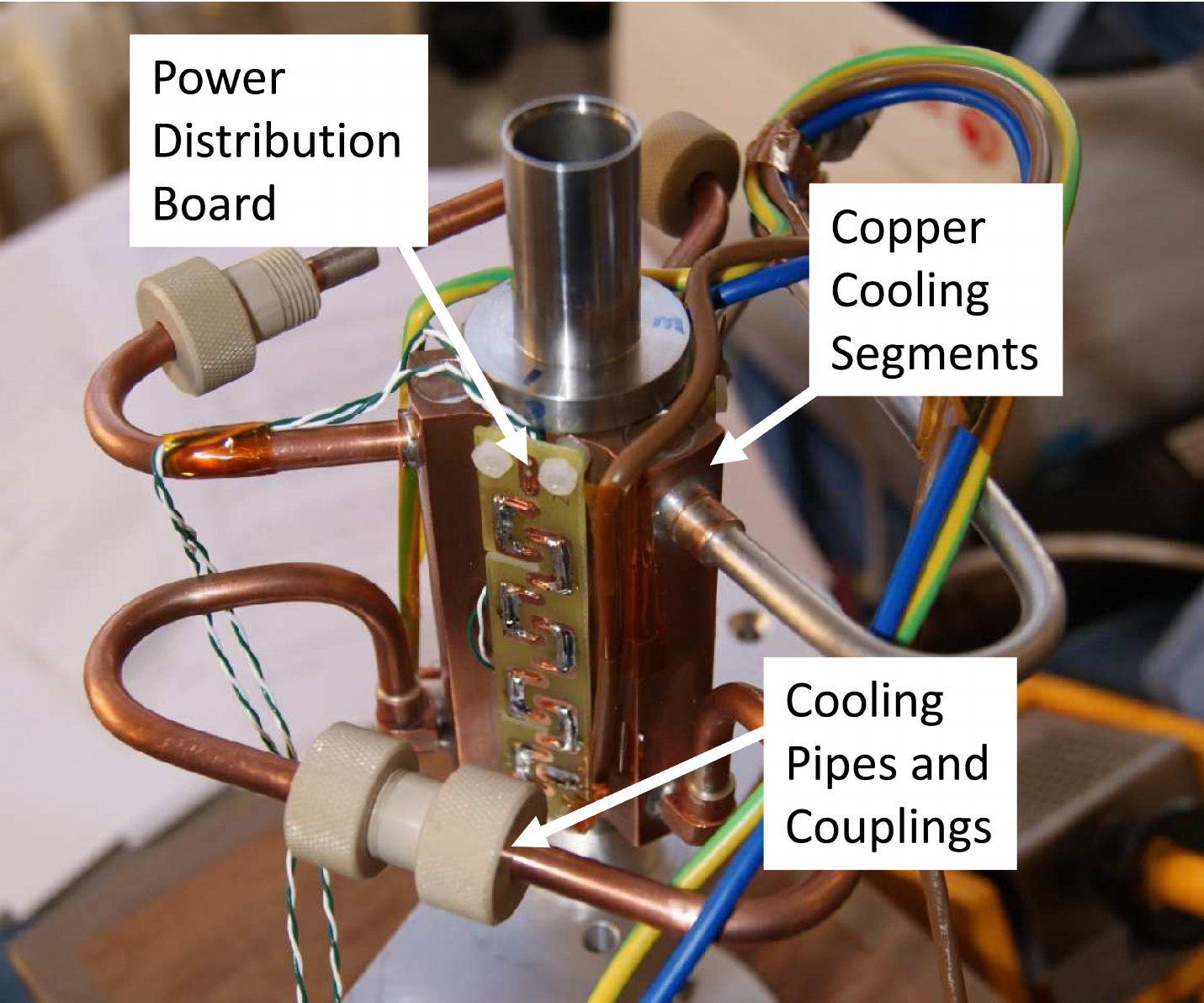}\\
\end{center}
  \caption{The stator with cooling jacket, before potting.}
  \label{fig:stator:cooling}
\end{figure}

The next step was to fit the water cooling jacket. 
An improvement in cooling means that the coils can be operated at
a higher current, to obtain more force on the permanent magnets mounted on
the target shaft and so higher acceleration.
Alternatively, for the original current the temperature of the coils is 
reduced, their resistance is lower and ohmic heating of the assembly is 
reduced.
The previous cooling jacket consisted of a coiled copper pipe brazed to
a copper sheet wrapped around the stator.
Improvements have been made to both the thermal contact of the jacket with
the coils and coolant flow rate through the jacket.
The new cooling jacket is formed from three copper segments which are 
clamped to the outer diameter of the bobbin and then 
plumbed together in series to allow one water input and one output from 
the assembly. 
Once the cooling jacket was fitted, a potting jig was used to allow the 
complete assembly to be potted under vacuum in such a way as to allow it to 
fit inside the target body. 
The potting not only ensures that the stator is robust; it also improves the 
heat path to the cooling jacket and provides a further electrical insulation 
barrier.
The heat transfer to the cooling segments was further enhanced by the choice 
of potting compound used.
This incorporated nano-spheres, which both increased the thermal conductivity 
and the fluidity of the compound.
The stator, complete with power distribution board, cooling jacket and pipes,
is shown in figure \ref{fig:stator:cooling} prior to potting.
Tests conducted on the new stator after potting 
showed a significant increase in cooling 
capacity over the previous design, as anticipated.

Once potted, the stator assembly was mounted inside the target body, which 
had been pre-machined to hold the stator assembly in place without axial 
movement.
The assembly was then electron-beam welded to ensure a vacuum seal for the 
ISIS beam line. 
The welding and vacuum testing took place at Electron Beam Services Ltd, 
Hemel Hempstead.  
The completed assembly was taken to Daresbury Laboratory for 
field mapping, as described in section \ref{Sect:Stator:Field}.

%
\subsection{Quality Control}
\label{Sect:Stator:QC}
%
The quality control for the stator consisted of mechanical checks on components
and electrical measurements as the device was assembled.
Once the mandrel, onto which the coils were wound, had been machined and
spark-eroded, it was surveyed by the Metrology Group at the 
Rutherford-Appleton Laboratory to ensure all dimensions were within tolerance.
After each coil was wound, the electrical insulation was tested at 1\,kV to ensure there 
was no breakdown between conductor and mandrel.
The measured resistance of each coil served to check for shorts from turn to 
turn or layer to layer, as did the measurement of inductance of individual
coils.
The inductance of a single coil was found to be $56.2 \pm 0.9\,\mu$H.
A final check of the electrical properties of the coils was provided by 
mapping the magnetic field, as described in the next section.

%
\subsection{Field mapping}
\label{Sect:Stator:Field}
%

\subsubsection{Procedure 1}
As indicated above, although target T1 performed well, subsequent targets 
did not run as smoothly 
and created small amounts of dust due to wear on the bearings. 
It was thought that the magnetic axis of the coils might not be coincident with 
the mechanical axis of the bearings or in the worst case the coils did not 
share the same axis, causing a sideways force on the bearing/shaft 
interface.

\begin{figure}
\begin{center}
\includegraphics[width=0.9\textwidth]{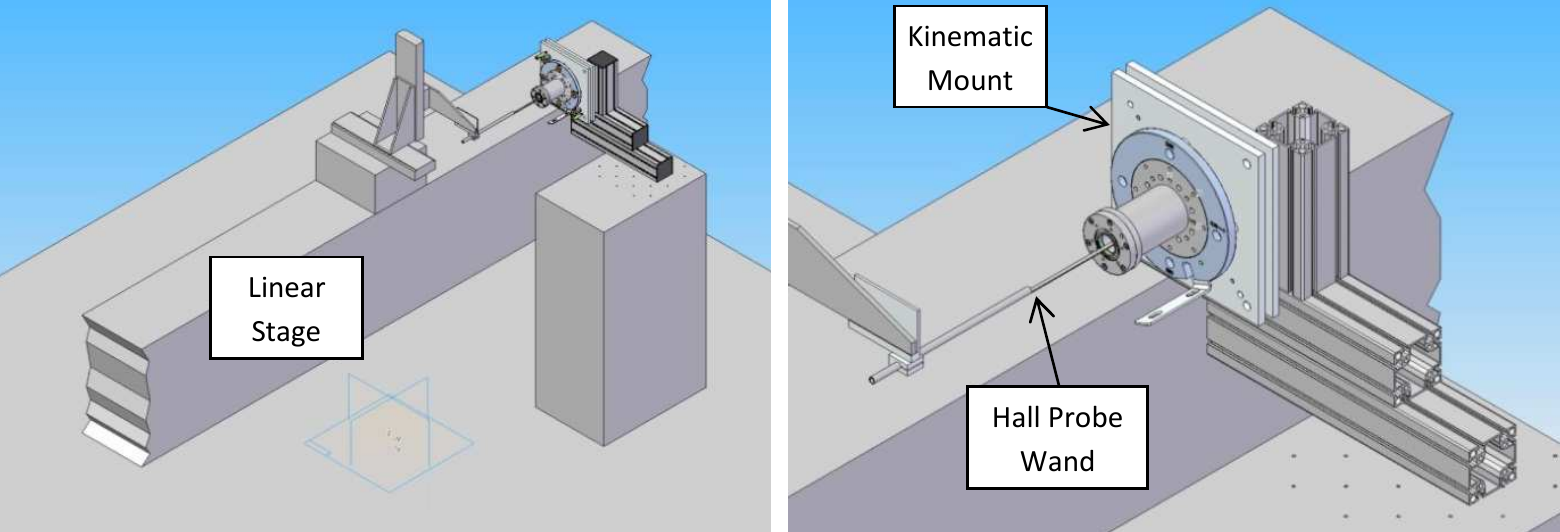}\\
\end{center}
  \caption{Stator field measurements at the Diamond magnet laboratory.}
  \label{fig:stator:diamond}
\end{figure}

The initial investigations took place at the Diamond magnet laboratory at the 
Rutherford Appleton Laboratory, with the apparatus illustrated in Figure 
\ref{fig:stator:diamond}.
A 3-axis Hall probe mounted on a `wand' was passed through the bore of the 
stator, its position controlled by linear stages. 
The stator was powered to around 6\,A. 
The mechanical axis was aligned to the linear stage; this allowed the Hall 
probe to pass along the stator at a known transverse offset, taking field 
readings at set points. 
This was done at different offsets and enabled a field map to be drawn up 
and analysed. 
This showed that the coils indeed did not share a common axis, i.e. they were 
transversely displaced with respect to one-another. 
It also showed that the mean field axis deviated from the mechanical axis.

Due to pressures on the Diamond magnet laboratory, the whole operation was moved to 
the STFC Daresbury Laboratory. 
The procedure remained the same -- a lengthy set-up and alignment period using a 
surveying telescope and adjustment of the stator body on a kinematic mount to 
align the stator bore with the long axis of the linear stages. 
The T series targets were all mapped in this way, and it became apparent that 
the longest time was taken in the set-up. 
This was also the most likely time to introduce errors so an alternative 
procedure was required.
The above measurements are detailed in the previous paper on the MICE target
\cite{PaperI}. 
They were able to show that there was a transverse offset between the coils and 
the mechanical axis, but were only able to measure the offset to an accuracy 
of around 100\,$\mu$m due to alignment limitations.

\subsubsection{Procedure 2}
The mapping lab at Daresbury has a rotation ($\theta$) stage. 
To avoid the long tedious set-up we embedded a single-axis Hall probe into a 
brass shaft 15\,mm in diameter, with the Hall probe offset from the centre and 
aligned to measure the longitudinal field component. 
This assembly slid through bearings mounted in the same tapered bearing-mounts 
that housed the bearings used when the target was in operation.
This meant that without any set-up time the Hall probe was aligned with
the mechanical axis. 
We also used the $\theta$ stage to rotate the shaft probe-assembly. 
This allowed measurements to be taken at different $\theta$ positions while 
ensuring the same radial distance from the mechanical axis. 

\begin{figure}
\begin{center}
\includegraphics[width=0.9\textwidth]{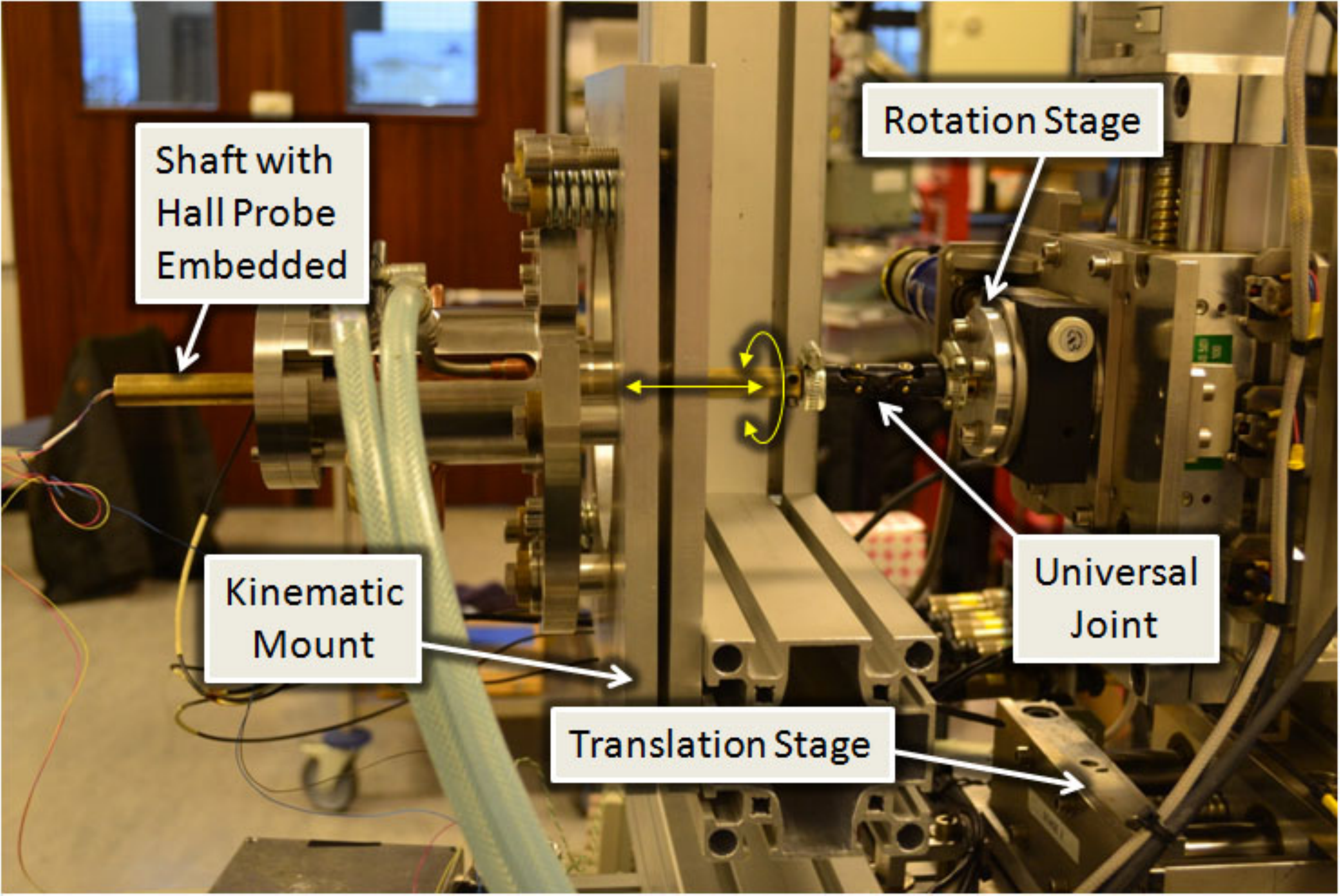}\\
\end{center}
  \caption{The revised stator measurement rig at Daresbury Laboratory. 
The Hall probe shaft is shown inserted into the stator. 
The directions of movement are indicated with yellow arrows.}
  \label{fig:stator:revised}
\end{figure}

Set-up consisted of simply mounting the stator on the kinematic mount and 
feeding the Hall probe shaft through the bearings. 
It then engaged into a double universal joint which meant that accurate 
alignment was not required. 
Figure \ref{fig:stator:revised} shows a picture of the measurement set-up. 

Field mapping was carried out in two stages. 
Firstly, the stator's longitudinal field profile ($B_{z}$ versus $z$) was measured. 
Secondly, the probe was moved to each peak in turn, and the shaft was rotated 
around the stator axis in order to measure the transverse field profile ($B_{z}$
versus $\theta$) for each coil.

Analysis of the field data consisted of fitting a pair of sinusoid functions to 
the data:
\[B_{z} (\theta)=B_0 + B_1 \sin(\theta +\phi_1 )+B_2 \sin(2\theta + \phi_2 )\;. \]
This fit was generally found to be very accurate, with very low rms residuals 
(typically less than 10\,$\mu$V, on a Hall-probe voltage level of around 600\,$\mu$V). 
The $B_1$ term is related directly to the coil offset, $d$. 
This was modelled using Opera-3D to give an approximate formula linking coil 
offset to the magnitude of the first sinusoid:
\[ \frac{B_1}{B_0} = a_2 d^2 + a_1 d + a_0\;. \]
The coefficients $a_2$, $a_1$, and $a_0$ were found to be 0.0166\,mm$^{-2}$, 
0.0782\,mm$^{-1}$ and $-2.7 \times 10^{-6}$ respectively.
The $B_2$ term is a `quadrupole-like' term and is thought to arise from a 
deformation of the coil. 
A coil with a highly elliptical shape would have a large $B_2$ term. 
In all the measured data, the $B_2$ term was not large enough to cause concern.

The newest version of the stator (S3) has an additional electrical 
connection -- the `star' connector. 
This allows individual phases (i.e. each set of eight coil -- phases A, B, C) 
to be connected up separately, and is of great help for measurements since it 
allows the field from individual coils to be distinguished more clearly.

\begin{figure}
\begin{center}
\includegraphics[width=0.9\textwidth]{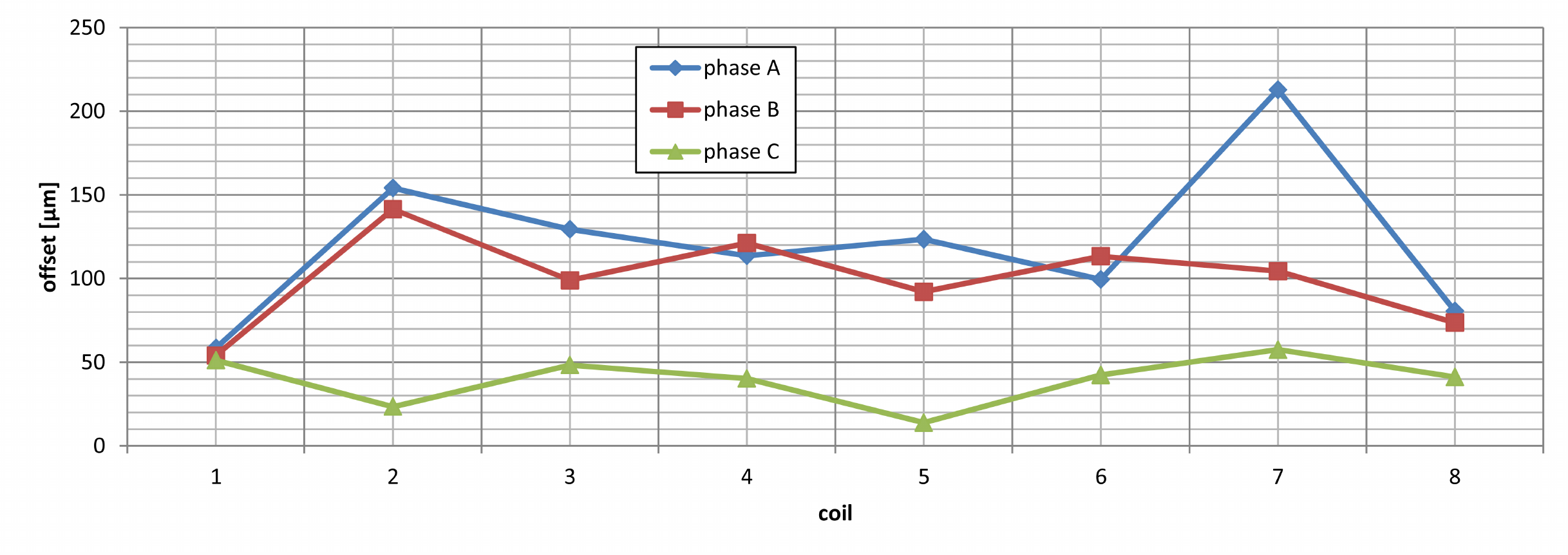}\\
\end{center}
  \caption{The measured offsets of each coil for stator S3.}
  \label{fig:stator:offsetsS3}
\end{figure}

\begin{figure}
\begin{center}
\includegraphics[width=0.9\textwidth]{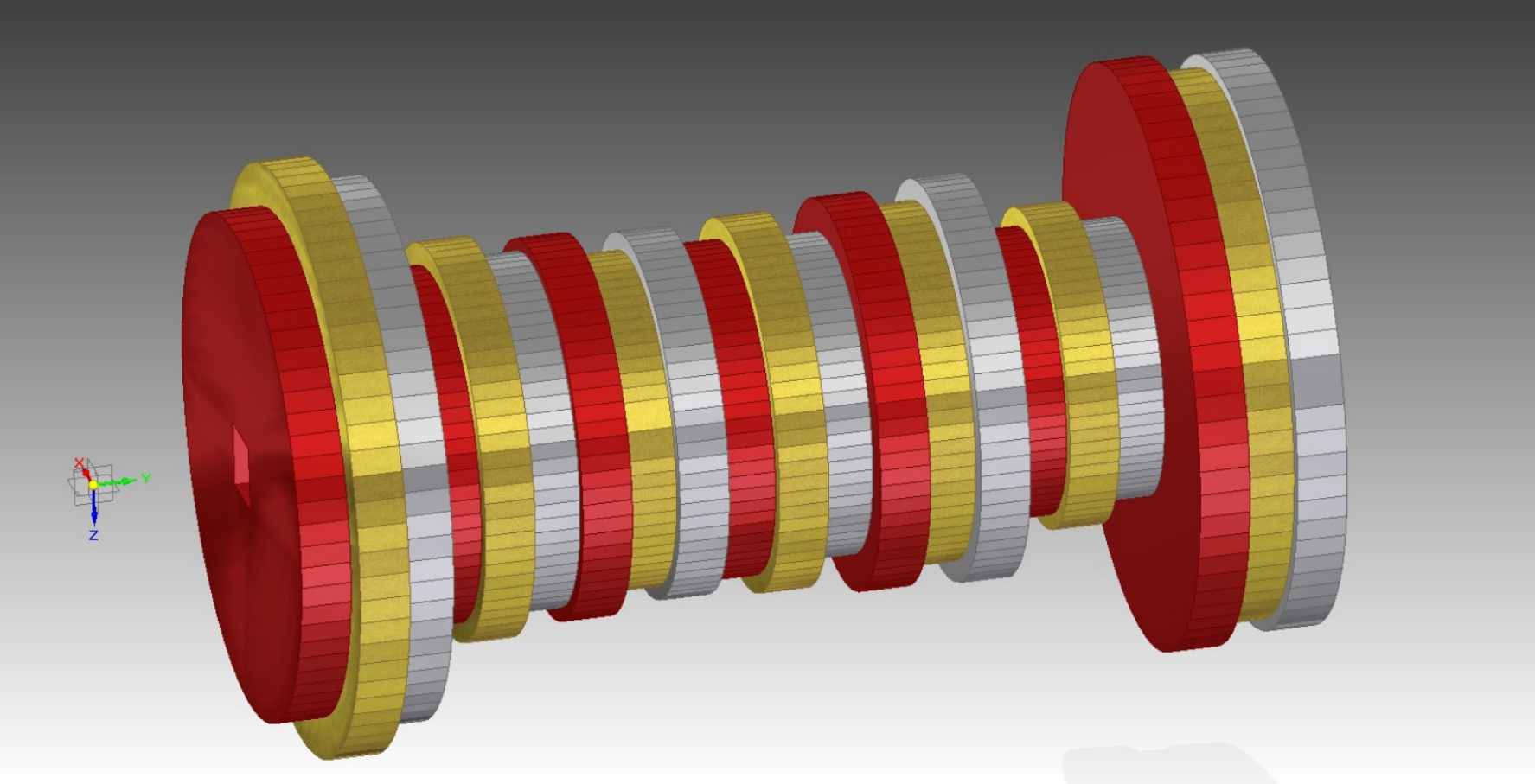}\\
\end{center}
  \caption{Graphic representation of the field map.
The three phases (A\,=\,red, B\,=\,gold and C\,=\,silver) show slight offsets
which are 120$^{\circ}$ apart.}
  \label{fig:stator:graphic}
\end{figure}

The measured coil offsets for stator S3 are shown in Figure \ref{fig:stator:offsetsS3}. 
The average coil offset is 87\,$\mu$m, with a standard deviation of 41\,$\mu$m.
An alternative representation of the field-map data is provided in Figure
\ref{fig:stator:graphic}.
Here the angle and magnitude of the Hall-probe output is used as a vector
to create a series of space points that form a circle.
These are turned into solid discs and shown assembled on the common bearing
axis.
It is interesting to note that the three phases have slight offsets which are 
120$^{\circ}$ apart.
This is believed to be due to the alignment of the step up from layer to layer 
in each coil winding.

Measurements were also taken without using the star connector -- in other words 
connecting two phases in series (A-B, A-C, B-C). 
This was done to provide a comparison with previous stators that did not have 
a star connector. 
The previous stator, S1, was remeasured using this technique to provide a 
comparison.
The stators showed similar results (Table \ref{tab:stator:comparison}).

\begin{table}
\begin{center}
    \begin{tabular}{c c c}
    \hline
    {\bf stator}&{\bf Mean coil offset}&{\bf Coil offset standard deviation} \\
    \hline
    {\bf S1}&79\,$\mu$m&30\,$\mu$m\\
    {\bf S3}&83\,$\mu$m&41\,$\mu$m\\
    \hline
    \end{tabular}
\end{center}
\caption{Comparison of stators S1 and S3.}
\label{tab:stator:comparison}
\end{table}


%
\section{Controller}
\label{Sect:Controller}

\begin{figure}
\begin{center}
\includegraphics[width=0.9\textwidth]{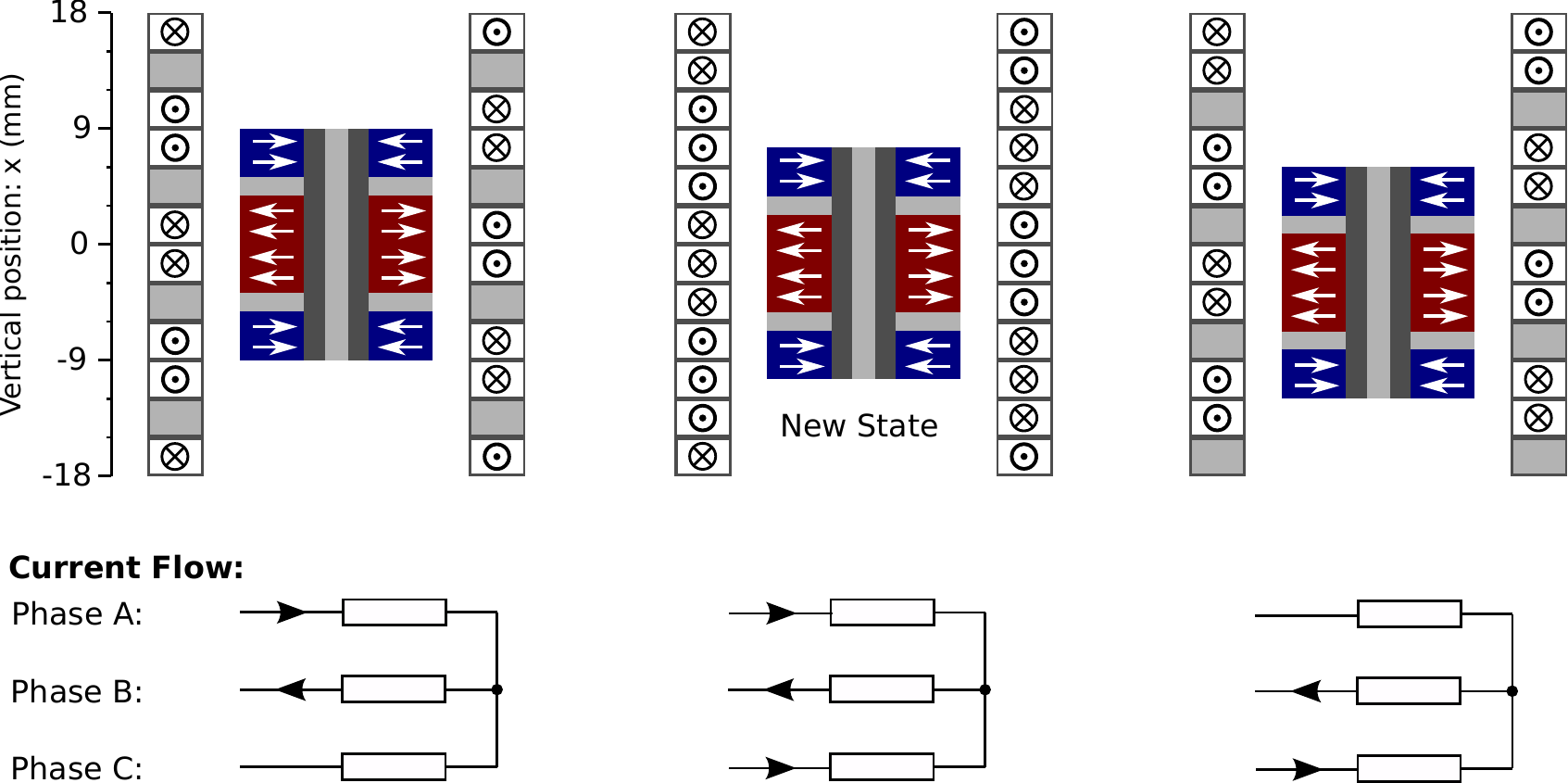}\\
\end{center}
  \caption{Motor drive states, with dots and crosses indicating current 
flow out of and into the page, while the radial polarisation of the permanent 
magnets is indicated by arrows.  
The permanent magnets are located to deliver the maximum positive acceleration. 
The new state occurs between two existing states and utilises all 
three sets of coils.}
  \label{fig:Controller:States}
\end{figure}

In addition to hardware improvements, an improved drive method has been 
implemented in the target-controller's firmware. 
The stator is configured with three sets of coils each wired in series.
These coils are joined at a single star point. 
As described in \cite{PaperI}, the motor has previously been driven by powering two sets of coils at any one time, so that current flows in through one set and out through the other; the third set of coils was unpowered.
To increase performance of the mechanism a total of six new drive states have 
been added, which power all three sets of coils simultaneously, and these are 
each located between two existing states. 
Figure \ref{fig:Controller:States} shows a side view of the coils in one of the 
new drive states.

The new state effectively powers two sets of coils in parallel, while the 
third is driven in series. 
This reduces the effective resistance of the stator by a factor of
three-quarters and so increases the power deposited in the stator by a factor 
of four-thirds, since the stator is driven from a fixed voltage source.

During actuation the new drive states are active one half of the time, which 
leads to an increase in power deposited within the stator by one sixth and 
increases the average acceleration of the mechanism by 17\%. 
In addition each of the new drive states adds an extra zero-force point at 
which the target can be levitated, which doubles the granularity of the motor 
and has levitation points separated by 1.5\,mm. 
The improved granularity reduces the maximum jitter at both the starting and 
minimum points of the actuation, which in turn reduces variability in the 
beam-loss generated in the synchrotron.

%
\section{User interface}
\label{Sect:GUI}


\subsection{Overview of the control software}
The user interface allowing control of the target has been completely redesigned
to allow a much more flexible and maintainable code base. 
The various target controls (stop, start, raise, lower, 
etc.) are accessed via a web interface using any standard web browser. 
A target control PC both runs the low level target control code written in C++ 
and also acts as a web-server handling HTTP requests from any authorised 
computer. 
The user, usually a non-target-expert MICE operator, needs to be able to steer 
the target into the ISIS beam and monitor the consequent beam loss, moving the 
target as required to accommodate changing requirements from the shift leader 
and run plans.

\begin{figure}
\begin{center}
\includegraphics[width=0.9\textwidth]{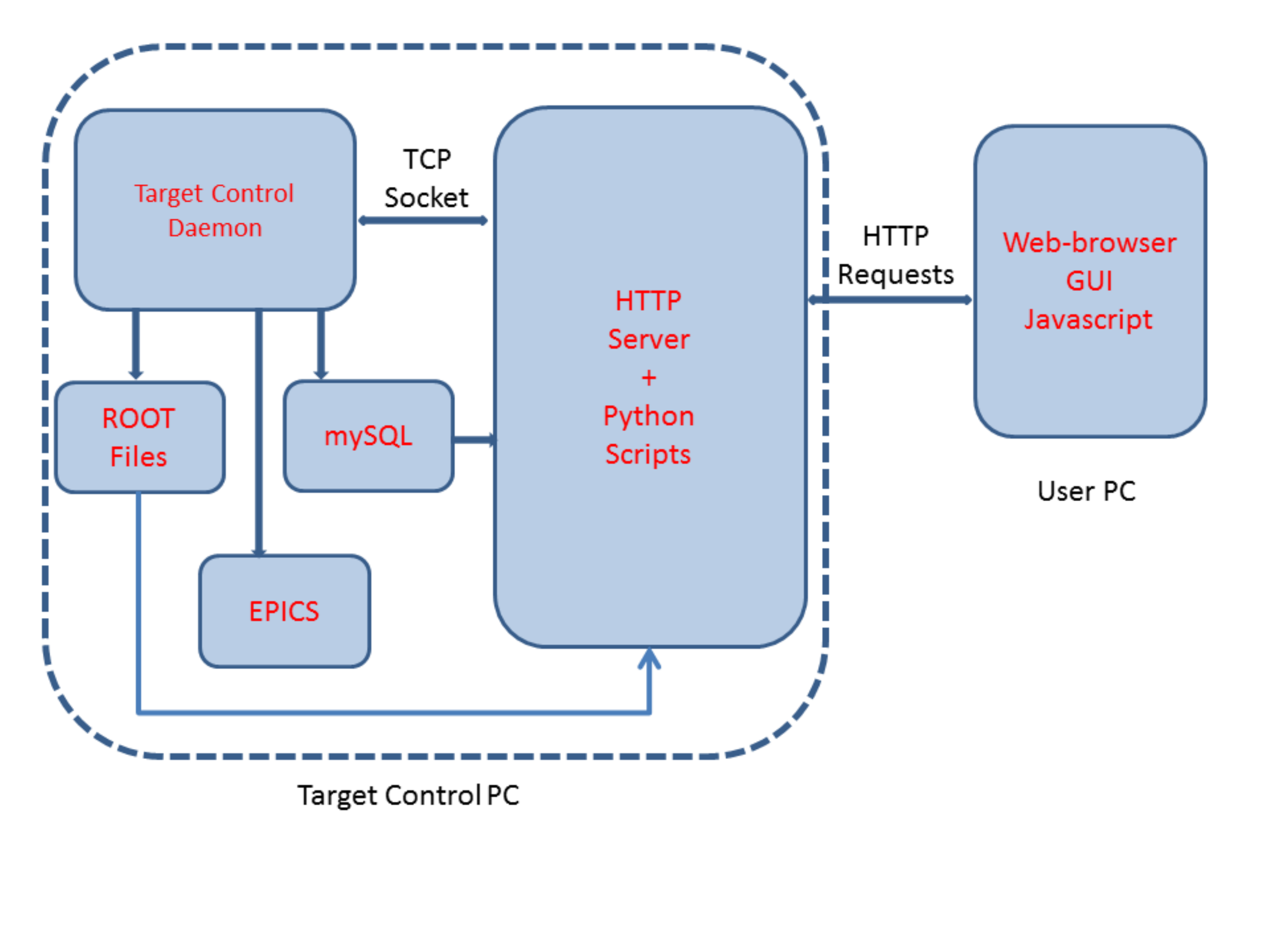}\\
\vspace {-15 mm}
\end{center}
  \caption{The MICE Target Control System.}
  \label{fig:GUI:Diagram}
\end{figure}

The target control PC runs the control daemon with responsibility for 
communicating with the FPGA target controller via a USB link. 
The daemon also collects data from the position digitiser and writes this data to 
disk. 
Configuration settings and target actuation parameters are read from and 
stored in a MySQL database. 
A subset of the actuation parameters are output as EPICS\cite{epics} process variables 
allowing interaction with the MICE Controls and Monitoring software. 
A custom set of ROOT\cite{root} scripts runs over the stored data in real time and produces
a set of ROOT files containing the full set of parameters (time, Beam Centre 
Distance, etc.) 
allowing the reconstruction of the complete target trajectory for each 
actuation. 
The control PC also acts as an HTTP web-server allowing the use of a 
web-browser-based control system. 
The web-server handles HTTP requests and displays the current target status 
including the full trajectory, together with other key parameters. 
Actuation history can be referenced using the same interface, in which case the 
web server pulls the requested data from the database. 
User input to control the target is sent over the same HTTP connection and is 
relayed to the control daemon process.
The communication flow is shown in Figure~\ref{fig:GUI:Diagram}.

\subsection{The Target Control Web Interface}
\begin{figure}
\begin{center}
\includegraphics[width=0.9\textwidth]{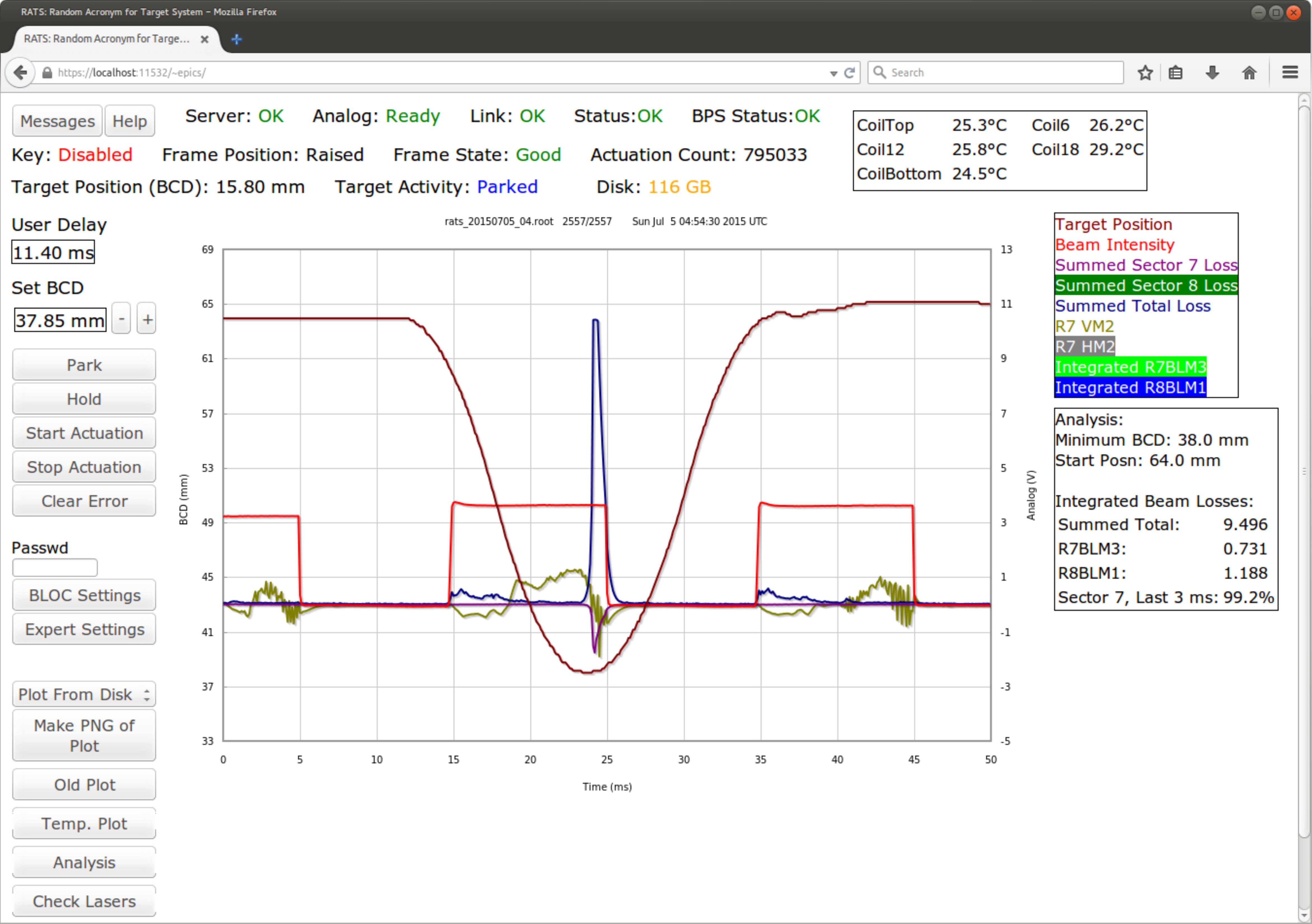}\\
\end{center}
  \caption{The Target Control Interface.}
  \label{fig:GUI:GUI}
\end{figure}

The target control web interface (shown in Figure \ref{fig:GUI:GUI}) is a 
JavaScript-generated web 
form displaying the key target operational parameters.

The upper section of the form displays basic status information including the 
target position, actuation status, actuation count and key interlocks. 
The left hand portion of the screen allows the user to set the target timing 
delay (the delay in milliseconds from a trigger based on the synchrotron
injection) and the minimum
beam centre distance, BCD. 
These are the two key parameters allowing the steering of the target into an 
optimal position to produce the required particle rate while not exceeding
the allowed ISIS beam loss.  
A further set of button controls allow the user to stop and start the target 
actuation and place it into park or hold states. 
The bottom set of control buttons allow access to expert settings and are 
password-protected, limiting their use to target experts. 
The main central section of the interface shows a customisable display of the 
target trajectory where the specific plots required are selectable via the box 
to the right. 
Usually the user will want to view the target position together with the 
beam intensity and one or more of the beam-loss values. 
Finally the box on the lower right of the GUI displays a set of key target 
parameters including the minimum BCD, the start position and integrated beam 
losses. 
The beam losses are particularly important as there is a maximum allowable 
beam loss set by ISIS and if this value is repeatedly exceeded MICE can trip 
the ISIS beam.

The use of a web-based control interface has many advantages. 
It greatly simplifies the deployment of the control software, which only exists 
on the web server and can easily be maintained and updated. 
Any MICE networked machine can connect to the target control PC via a simple 
URL and then be used to monitor and control the target. 
Security is simple to implement as only a specific named set of machines, based 
on their IP addresses, are allowed to connect. 
Also the layout of the web form GUI is relatively simple to implement as it is 
created using standard JavaScript and thus updates and modifications can
easily be accommodated.

%
\section{Performance}
\label{Sect:Performance}


A crucial element of the target system is the indirect 
monitoring of the mechanical performance, in order to
observe the 
wear profile of the mechanism.
From such observations requirements can be derived which
allow ``poor'' targets to be identified.
The mechanical performance is monitored through a change of
friction, which can only be observed indirectly.


\begin{figure}
\begin{center}
\includegraphics[width=0.9\textwidth]{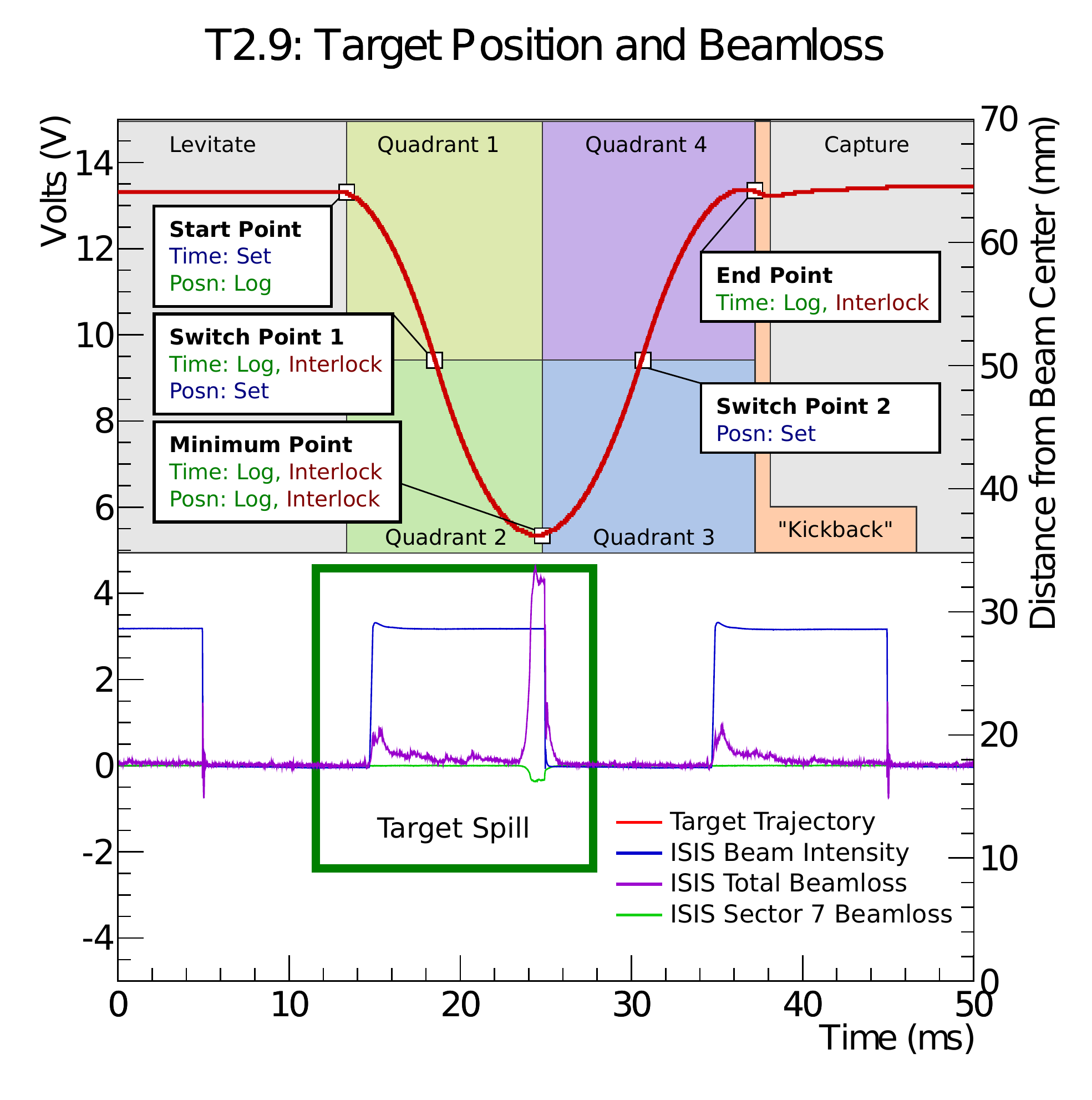}\\
\vspace {-10 mm}
\end{center}
  \caption[Annotated target trajectory]
  {Annotated target trajectory, showing the key actuation states and
  the transitions between. The ISIS beam intensity and losses are shown below.}
  \label{fig:perform:trjectory}
\end{figure}

In previous studies the acceleration change in the first
quadrant (see figure~\ref{fig:perform:trjectory}) and the starting-position 
width have been used to
monitor performance. An increase in friction leads to a
reduced acceleration in the first quadrant. Furthermore,
at the end of the actuation the target is captured in a
potential well and an increase in friction allows the 
target to stop further from the centre of the well,
which is observed as an increase in start-position width.
The acceleration change is strongly affected by external
factors, such as temperature, or driving voltage and the
start-position width is also affected by voltage and
controller configuration.

\subsection{Friction Measurement}
The most recent technique developed for monitoring the
performance makes a measurement of the average friction
observed during an actuation\cite{Overton}. This can be obtained
by considering the forces present in each
quadrant of the target's actuation, shown in
figure~\ref{fig:perform:trjectory}. 

In the first quadrant the driving force is downwards,
with friction in opposition, while the second quadrant
has friction assisting the driving force to bring the system
to a halt. In addition, the target is powered from a
capacitor bank, the discharge of which results in a reduced driving
force as a function of time in the actuation. The
observed acceleration in each quadrant of the actuation
is:
\begin{align}
a_1(t = t_\text{start}) &= -a_0  \exp \left(\frac{-t}{t_\text{RC}} \right) + \overline{F(x)} - g\;; \label{eqn:perform:aq1} \\
a_2(t = t_\text{min}) &= +a_0 \exp \left(\frac{-t}{t_\text{RC}} \right) + \overline{F(x)} - g\;; \\
a_3(t = t_\text{min}) &= +a_0 \exp \left(\frac{-t}{t_\text{RC}} \right) - \overline{F(x)} - g\;; \\
a_4(t = t_\text{end}) &= -a_0 \exp \left(\frac{-t}{t_\text{RC}} \right) - \overline{F(x)} - g\;; \label{eqn:perform:aq4}
\end{align}
where $a_0$ is the absolute maximum driving acceleration,
$t_\text{RC}$ is the decay time of the capacitor bank,
$\overline{F(x)}$ is the average friction and $g$ is the
acceleration due to gravity.
In order to avoid the need to consider velocity-dependent
characteristics, such as eddy currents, the acceleration
is determined at the stationary point on the border of each
quadrant.

To obtain the friction, first the acceleration of the
mechanism must be measured at each stationary point.
This is accomplished using
a third order polynomial fit, the second derivative
of which is evaluated at the stationary point.
Quadrants two and three share a common stationary
point, however each polynomial-fit range is over 
a different quadrant, hence a different value is 
obtained and used. The predicted values from
equations~\ref{eqn:perform:aq1} to \ref{eqn:perform:aq4}
are then fitted to the measured values using a
non-linear least squares fit.

\begin{figure}
\begin{center}
\includegraphics[width=0.99\textwidth]{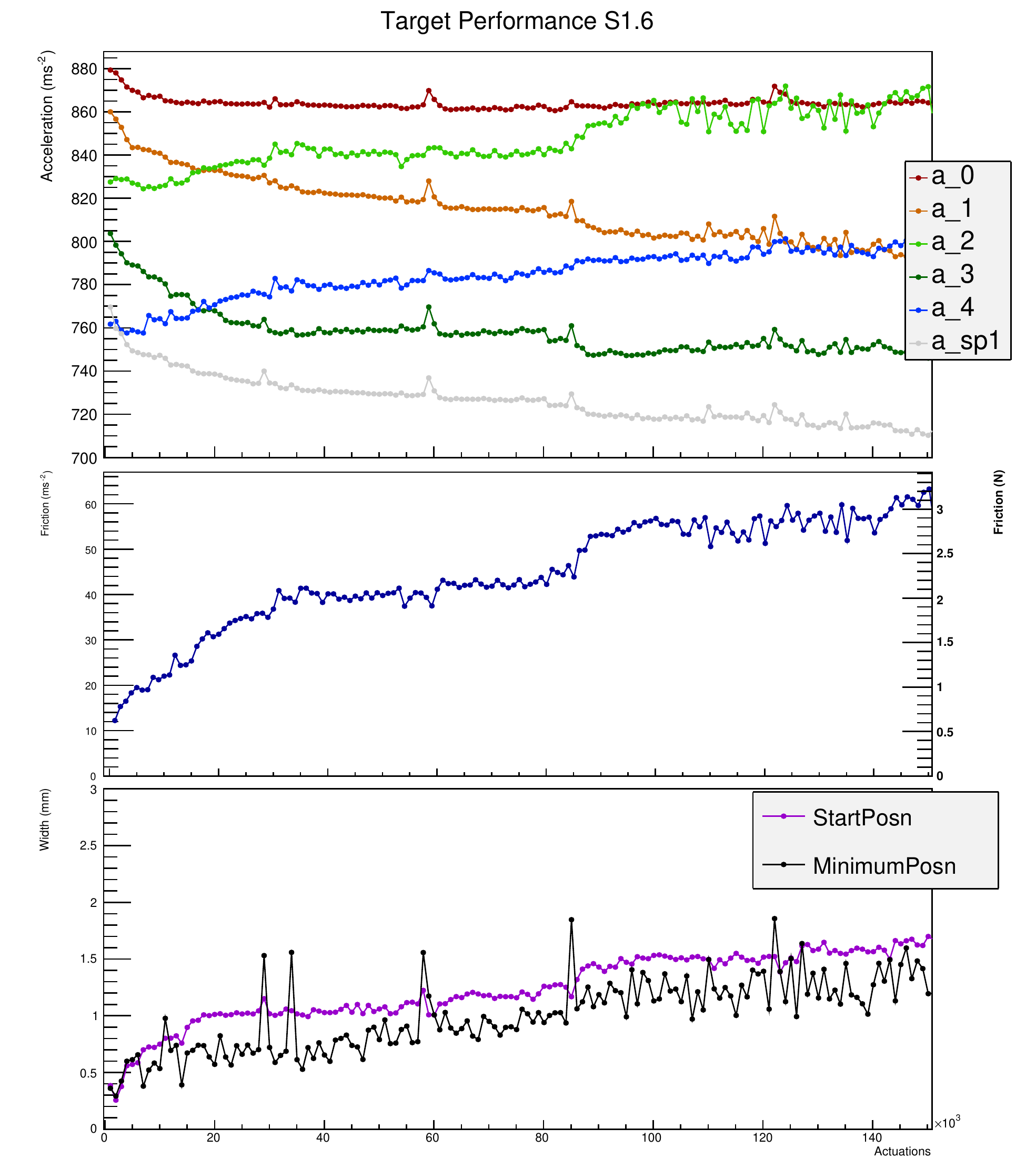}
\end{center}
  \caption[The acceleration magnitude, friction and start/minimum-position width for S1.6]
  {The acceleration magnitude, friction and start/minimum position width
  for target S1.6 during the first 150k actuations. The plotted value for
  both acceleration and friction is the median of 1000 samples recorded. The width
  is also measured over this range.
  Errors are not shown, but
  are dominated by systematic effects ($^{+2}_{-5} \text{ m s}^{-2}$).}
  \label{fig:perform:timeplot_S16}
\end{figure}

Results from this technique are shown in 
figure~\ref{fig:perform:timeplot_S16}, for the
first 150k actuations of target S1.6. 
The first plot
shows: the absolute values of the measured 
accelerations ($a_1$ to $a_4$), 
the fitted maximum acceleration ($a_0$) and the
average acceleration measured by the control
system in the first quadrant ($a_\text{SP1}$).
Over time the magnitude
of acceleration in quadrants 2 and 4 increased, while a decrease was observed
in quadrants 1 and 3. This change is interpreted as an increase in friction,
since in quadrants 2 and 4 the target is decelerating, which is shown in the
middle plot of the figure.
The final plot shows the measured width of both the starting and minimum positions;
as expected both show an increase as the measured friction increases.

\subsection{Wear Profile and Quality Assurance}
\begin{figure}
\begin{center}
\includegraphics[width=0.99\textwidth]{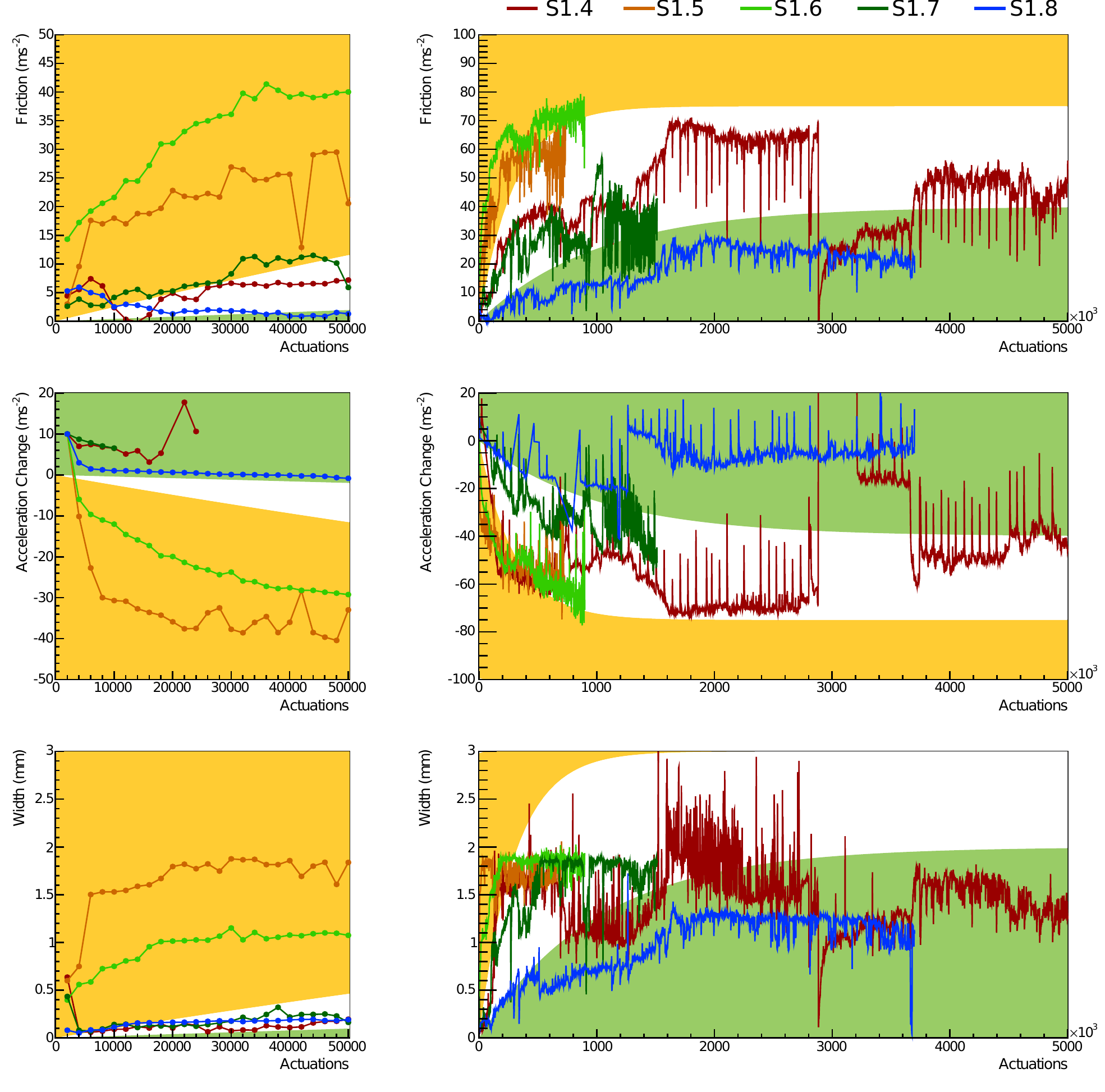}
\end{center}
  \caption[Performance evolution of S1.4 to S1.8]
  {
    QA (left) and normal running (right) of the S1 stator. The friction,
  acceleration change and start-position width are shown. The yellow
  band indicates the unacceptable performance region and the green band 
  identifies the ideal region.
  Missing 
  acceleration-change data points are due to an analysis cut which enabled
  only a single start-position to be used in measuring the acceleration.}
  \label{fig:perform:qa_s1_45678}      
\end{figure}

%
During each target test the performance of the mechanism is
logged, and this provides a guideline for the wearing of a typical
target. By considering the lifetime of the targets, regions
identifying poorly performing targets and well performing targets
can also be created, shown in figure~\ref{fig:perform:qa_s1_45678}.
The right-hand plot shows the performance over five million actuations;
as can be seen, targets S1.5 and S1.6 were unable to reach one million
and their lines terminate early. 
The left-hand plots show the performance during the first day
of running, which is used to assess performance and predict a target's
lifetime prior to installation on the accelerator. 
Both S1.5 and S1.6
fall within the poor-performance region and would be rejected before
being mounted on ISIS.

A summary of all target tests conducted is presented in 
table~\ref{tab:perform:stator_tests}.
12 targets in total have been studied, two of which failed
before reaching 1 million actuations.
Both the new S1 and original T1 stators have demonstrated that
targets accepted by the quality assurance process have operated for 
over 1.8 million
actuations. In addition, a number of targets have delivered
in excess of 5 million acceleration cycles. 

\begin{table}
\renewcommand{\arraystretch}{1.2}
\newcommand{\specialvalueCRtime}{$^{+0}_{-50}$}
\begin{center}
    \begin{tabular}{ l c c c c c }
    \hline
           & \multicolumn{4}{c}{50K tests}              & Number of  \\
    Target & Friction & Acceleration & Width & Accepted & Actuations (million)  \\
    \cmidrule(lr){1-1}\cmidrule(lr){2-5}\cmidrule(l){6-6}
    S1.8 & P & P & P & Y & 8.5+\\
    S1.7 & p & - & P & Y & 1.5\\
    S1.6 & F & F & F & N & 0.9\\
    S1.5 & F & F & F & N & 0.7\\
    S1.4 & P & - & P & Y & 10\\
    S1.3 & - & f & P & Y & 1.8\\
    S1.2 & - & P & P & Y & 3.1\\
    \cmidrule(lr){1-1}\cmidrule(lr){2-5}\cmidrule(l){6-6}
    T1.3 & - & f & P & Y & 5\\
    T1.2 & - & P & P & Y & 2\\
    T1.1 & - & P & P & Y & 3.2\\
    \cmidrule(lr){1-1}\cmidrule(lr){2-5}\cmidrule(l){6-6}
    T2.8 & - & F & P & Y & 2.5\\
    T2.7 & - & P & P & Y & 1.3\\
    T2.6 & - & P & P & Y & 1.1\\
    \hline
    \end{tabular}
\end{center}
\caption[Summary of stator QA results]
        {Summary of stator QA results, in date-descending order.
         P(p) = Pass(borderline). F(f) = Fail(borderline).
         1 million actuations corresponds to 14.8 days
         of continuous running at 50/64\,Hz (MS/64).
         }
\label{tab:perform:stator_tests}
\end{table}

%
\section{Summary}
\label{Sect:Summary}

The improved target mechanism for the Muon Ionisation Cooling Experiment has 
been presented.
A new technique for winding coils in situ, which also allowed smaller
coil-to-magnet clearance and more efficient water-cooling, has allowed a
more reliable and efficient linear motor to be employed.
A new magnetic-field mapping procedure has enhanced the quality control, and
demonstrated the advantages of the new coil-winding approach.
An upgrade of the motor controller allows better exploitation of the 
hardware, with increased acceleration and more precise control.
In addition, a rewritten user interface provides easier use of the
target actuator within the MICE experiment and provides better
monitoring of its behaviour.
Finally, a new method of analysing performance data allows changes in friction
to be monitored and this has allowed poorly-performing bearing sets to
be identified before targets are installed on ISIS, as well as giving 
an early warning
of changes that indicate a target is reaching the end of its useful life.

%
%
\section*{Acknowledgements}

We gratefully acknowledge the ISIS Division at the STFC Rutherford
Appleton Laboratory for the warm spirit of collaboration  and for
providing access to laboratory space, facilities, and invaluable
support. 
We are indebted to the MICE collaboration, which has provided the
motivation for, and the context within which, the work reported here was
carried out.
We would like to acknowledge the contributions of
AC Precision, Standlake,
Electron Beam Services Ltd, Hemel Hempstead,
the workshop of Oxford University Physics Department,
Imperial College London's HEP electronics lab,
Dr.\,E.\,Longhi of the Diamond magnet laboratory
and the Metrology Group of Rutherford Appleton Laboratory.


This work was supported by the Science and Technology Facilities Council
under grant numbers
ST/H001735/1,ST/J002046/1, ST/K001337/1, ST/K00610X/1,
and through SLAs with STFC-supported laboratories.

%
\clearpage
\bibliographystyle{JHEP}

\bibliography{TargetPaperII}
%
\end{document}